\documentclass[12pt, reqno]{amsart}

\usepackage{amssymb, mathrsfs}
\usepackage{latexsym}
\usepackage{amsmath, commath}
\usepackage{amsthm}
\usepackage{amsfonts}
\usepackage{dsfont}
\usepackage{mathtools}
\usepackage{epsfig}
\usepackage{verbatim}
\usepackage{graphicx}
\usepackage{accents}

\usepackage{stmaryrd}

\usepackage{pstricks-add}

\usepackage{tikz}
\usetikzlibrary{cd}
\usetikzlibrary{patterns}
\usetikzlibrary{calc}
\usepackage{pgfplots}
\pgfplotsset{compat=1.11}

\usepackage[left=1.4in,right=1.4in]{geometry}
\usepackage{enumerate}
\usepackage[colorlinks=true,linkcolor=blue,citecolor=blue]{hyperref}

\usepackage{soul}

\usepackage{todonotes}

\usepackage[titletoc, title]{appendix}

\usepackage{bm}
\usepackage{bbm}
\usepackage{nicefrac}

\usepackage{slashed}

\newtheorem*{lemma*}{Lemma}

\newtheorem{theorem}{Theorem}[section]

\newtheorem{lemma}[theorem]{Lemma}

\newtheorem*{open question}{Open Question}

\newtheorem*{question*}{Question}

\theoremstyle{definition}

\newtheorem*{claim*}{Claim}
\newtheorem{definition}[theorem]{Definition}
\newtheorem{example}[theorem]{Example}

\theoremstyle{remark}

\numberwithin{equation}{section}

\begin{document}
\title[Quantum Complexity of Permutations]{Quantum Complexity of Permutations}

\author{Andrew Yu}
\address[]{ Harvard University, Cambridge, MA 02138,  and Phillips Academy, Andover, MA 01810, USA.}
\email{andrewyu45@gmail.com }
\thanks{Supported by ARO MURI grant W911NF-20-1-0082}

\maketitle

	\begin{abstract} Quantum complexity of a unitary measures  the runtime of quantum computers. In this article, we study the complexity of a special type of unitaries, permutations. Let   $S_n$ be the symmetric  group of all permutations of  $\{1, \cdots, n\}$ with two generators: the transposition 
and the cyclic permutation (denoted by $\sigma$ and $\tau$). The permutations $\{\sigma, \tau, \tau^{-1}\}$ serve as logic gates. We give an explicit construction of  permutations in $S_n$
	with quadratic quantum complexity lower bound $\frac{n^2-2n-7}{4}$. We also prove that all permutations in $S_n$ have quadratic  quantum complexity upper bound $3 (n-1)^2$. Finally, we show that almost all permutations in $S_n$ have quadratic quantum complexity lower bound when $n\rightarrow \infty$. The method described in this paper may shed light on the  complexity problem for general unitary in quantum computation.
	\end{abstract}
	
	$$\\$$

\section{Introduction}

$$\\$$

A quantum computation is a unitary transformation  implemented by a given finite set  of  unitary transformations called logic gates.
Quantum complexity of a unitary transformation 
 is then defined as  the smallest
number of logic  gates needed to implement the  unitary transformation (c.f. [1, 2, 3, 5, 6, 7, 8, 9, 10, 11, 12]). 
A central problem in quantum computation is  to estimate the quantum complexity of a unitary. In [1, 2, 7, 8, 11], a geometric method was developed to investigate quantum complexity.
In this article, we use a combination of geometric ideas with combinatorial techniques to study the quantum complexity of   permutations, a special type of  unitary transformations.  Our results can be applied to high energy physics and quantum field theory.

The main results of this article are as follows. First, we  give an explicit construction of permutations with quadratic quantum complexity lower bound $\frac{n^2-2n-7}{4}$.  Secondly, we show that all permutations have 
quadratic  quantum complexity upper bound $3 (n-1)^2$. Finally, we prove that almost all permutations have quadratic quantum complexity lower bound $\frac{1}{32} n^2-3$.


We define the symmetric group $S_n$ to be the group of all permutations of the set $\{1, \cdots, n\}$. Let $\sigma$ be the permutation:
$\sigma(1)=2, \sigma(2)=1$, and $\sigma(i)=i$ if $i\neq 1, 2$, 
and let $\tau$ be the cyclic permutation: $\tau(1)=2, \tau(2)=3, \cdots, \tau(n-1)=n$, $\tau(n)=1$. The permutations $\sigma$ and $\tau$ generate
the symmetric group $S_n$ in the sense that every element in $S_n$ can be written
as the product of a sequence of permutations from 
$\{\sigma, \tau, \tau^{-1} \}$  
(c.f. Theorem  1.4).

A quantum computation can be described as a sequence of logic gates. 
In our model, we  choose the generators $\sigma$, $\tau$, and  $\tau^{-1}$   as logic  gates. Since any permutation can be expressed as a product of a sequence of the  chosen logic gates,  a quantum computation is  equivalent to a permutation.  Quantum complexity of a permutation is then the smallest number in  a sequence of these gates needed to build the permutation.
It is a natural problem to estimate quantum complexity of permutations in $S_n$. 

$$\\$$

In the following theorem, we   construct an explicit  permutation with quadratic  quantum complexity lower bound  (such a permutation is also called quadratically  hard to implement).

\begin{theorem} If $\omega$ is the permutation of the set $\{1, 2, \cdots, n\}$ defined as follows:
$$\omega(1)=1, \omega(2)=3, \cdots, \omega (\frac{n+1}{2})  =n,  \omega (\frac{n+1}{2}+1)=2, \omega(\frac{n+1}{2}+2)=4, $$ $$\cdots, \omega(n)=n-1$$
when $n$ is odd, and 
$$\omega(1)=1, \omega(2)=3, \cdots, \omega(\frac{n}{2})=n-1, 
\omega(\frac{n}{2}+1)=2, \omega(\frac{n}{2}+2)=4,\cdots,  \omega(n)=n $$
when $n$ is even, then
 $\omega$
 has  quadratic  quantum complexity lower bound $\frac{n^2-2n-7}{4}$,
more precisely,  if 
 we write $$\omega=\rho_1 \cdots\rho_l$$ with each $\rho_i$ $(1\leq i\leq l)$ being a permutation from the generating set $\{\sigma, \tau, \tau^{-1}\}$, then $$l \geq \frac{n^2-2n-7}{4}.$$

\end{theorem} 

$$\\$$


The following very bumpiness concept provides the intuition why the permutation in Theorem 1.1 has quadratic  quantum complexity lower bound.


\begin{definition} We identify the set $\{1, 2, \cdots, n\}$ with the set of all integers modulo $n$.   A permutation $\omega$ of  $\{1, 2, \cdots, n\}$
is said to be very bumpy if, for each $k\in \{0, 1,2 \cdots, n-1\},$
$$\# \{\quad i \in \{1, 2, \cdots, n\}: \quad d(k+ \omega(i), i )\geq \frac{n}{8}\quad \}
\geq    \frac{n}{4} ,$$
where addition is performed  modulo $n$ and
 $d(x, y)= \min \{|x-y|, n-|x-y|\}$.

\end{definition}

$$\\$$

The next theorem gives a quadratic quantum complexity lower bound for very bumpy permutations. 

\begin{theorem} If a permutation $\omega$ of  $\{1, 2, \cdots, n\}$ is
 very bumpy, then $\omega$ has quadratic quantum complexity lower bound,
more precisely, 
 if 
 we write $$\omega=\rho_1 \cdots\rho_l$$ with each $\rho_i$ $(1\leq i\leq l)$ being a permutation from the generating set $\{\sigma, \tau, \tau^{-1}\}$, then $$l\geq \frac{n^2}{32}-3.$$
\end{theorem}


$$\\$$

The following result gives a quadratic upper bound for quantum complexity
of all permutations.

\begin{theorem} 
Every permutation in $S_n$ has quantum complexity upper bound $3(n-1)^2$, more precisely,
for any permutation $\omega \in S_n$, we can write 
$$\omega=\rho_1 \cdots\rho_j$$
with each $\rho_i$ $(1\leq i\leq j)$ being a permutation from the generating set $\{\sigma,  \tau, \tau^{-1}\}$ and 
$$j\leq 3 (n-1)^2.$$

\end{theorem}

Theorem 1.1 implies that the quadratic quantum complexity upper bound in the above theorem is optimal for permutations.
$$\\$$

Our final theorem says that a  permutation has quadratic quantum complexity lower bound with probability $1$ (as $n\rightarrow \infty$).

\begin{theorem} Almost all permutations have quadratic quantum complexity lower bound, more precisely, the ratio of the number of permutations in $S_n$ with quantum complexity lower bound $\frac{1}{32} n^2-3$ over $\# S_n=n!$ goes to $1$ when $n\rightarrow \infty$, where  $\# S_n$ is the number of elements in the symmetric group $S_n$.

\end{theorem}

We prove the above theorem by showing that almost all permutations are very bumpy.

$$\\$$

The rest of this article is structured as follows. In  Section 2,
we introduce a discrete geometry method to study quantum complexity lower bound. 
In Section 3, 
we give an explicit construction of permutations with  quadratic  quantum complexity lower bound and prove that very bumpy permutations have  quadratic  quantum complexity lower bound. 
In Section 4, we show that all permutations have  quadratic quantum complexity upper bound.
In Section 5, we prove that almost all permutations have quadratic quantum complexity lower bound.

$$\\$$

\section{Discrete geometry and  quantum complexity lower bound}\label{sec:pre}

$$\\$$

In this section, we  introduce a discrete geometry method for deriving  lower bounds for the quantum complexity of any permutation. In the next section, we will combine this method with other  combinatorial ideas to give an explicit construction of permutations with quadratic quantum complexity lower bound.

A pair of elements $\{ x, y\}$ in $\{1, 2, \cdots, n\}$ are said to be neighbors if $|x-y|=1$ or $\{ x, y\}=\{1, n\}$.
We will introduce a discrete path distance $d$ on the set $\{1, 2, \cdots, n\}$
such that the distance between two neighbors is $1$.

A function $\phi$ from $\{1, \cdots, k\}$ to $\{1, 2, \cdots, n\}$  is called a discrete path if $\phi(i)$ and $\phi(i+1)$ are either neighbors or  $\phi(i)=\phi(i+1)$ for each $i=1, \cdots, k-1$. The length of the discrete path 
$\phi$  is defined to be 
$$d(\phi(1), \phi(2))+ \cdots +d(\phi(k-1), \phi(k)),$$ 
where $$d(\phi(i), \phi(i+1))=1$$
 if $\phi(i)$ and $\phi(i+1)$ are  neighbors as defined above,
 and $$d(\phi(i), \phi(i+1))=0$$ if $\phi(i)=\phi(i+1)$.

A discrete path $\phi$ as defined above  is said to be connecting a pair of elements
$\{x,y\}$ in  $\{1, 2, \cdots, n\}$ if $$\phi(1) =x, ~~~~\phi(k)= y.$$

$$\\$$

\begin{definition} The discrete path distance on
$\{1, 2, \cdots, n\}$ is defined as follows.
For any pair $\{x, y\}$ in $\{1, 2, \cdots, n\}$, 
the discrete path distance between $x$ and $y$ is defined to be the length of the shortest
discrete path connecting $x$ to $y$.

\end{definition}

$$\\$$

The following basic fact plays an essential role in the proof of Theorem 1.1 and Theorem 1.2.

\begin{lemma} The discrete path distance between any pair of point $x$ and $y$ in  $\{1, 2, \cdots, n\}$ is equal to 
 $$d(x,y)=\min\{ |x-y|, n-|x-y|\}.$$
\end{lemma}

$$\\$$

\noindent{Proof of Lemma 2.2. } 
We place $\{1, 2, \cdots, n\}$ on a circle such that two neighbors from
$\{1, 2, \cdots, n\}$ are adjacent to each other on the circle.
One way to do this is to place each $k$  from $1$ to $n$ at the position $exp(\frac{2\pi k i}{n})=cos(\frac{2\pi k i}{n}) +i ~sin(\frac{2\pi k i}{n})$ on the unit circle in the complex plane. 









 We claim that there exists a shortest discrete path $\phi$
 connecting $x$ to $y$ such that 
 
\noindent{(1)} $\phi(i)\neq \phi(i+1)$ for all $i=1, \cdots,k-1;$ 

\noindent{(2)} the sequence $\phi(1), \cdots, \phi(k)$, is ordered either clockwise or anti-clockwise on the circle.

 We first show that there exists a shortest discrete path $\phi$ such that $\phi(i)\neq \phi(i+1)$ for all $i=1, \cdots,k-1.$ Let $\phi$ be a shortest discrete path connecting $x$ to $y$. Assume by contradiction, there exists $i_0$ such that $\phi(i_0)= \phi(i_0+1)$. We can define a new discrete path $\phi'$ connecting $x$ to $y$ as follows: $\phi'(i)=\phi(i)$ for $i\leq i_0$ and $\phi'(i)=\phi(i+1)$
for $i_0+1\leq i\leq k-1$. Note that $\phi'$ is a discrete path defined on $\{1, \cdots, k-1\}$
satisfying $\phi'(1) =x$ and $\phi'(k-1)=y$.  It is easy to see that the length of $\phi'$ is equal
to the length of $\phi$.  If $\phi'$ satisfies the above condition (1), then $\phi'$ is our desired discrete path. Otherwise we repeat  the same  procedure until we  obtain a shortest path satisfying the above condition (1).
Notice that this process stops  at some point since each time we perform this procedure, the size of the domain of the new discrete path decreases by one element.

We now assume that $\phi$ is a shortest discrete path connecting $x$ to $y$ and
satisfies condition (1).
Without loss of generality, we can assume that $\phi(1)$, $\phi(2)$ are ordered 
clockwise on the circle. Assume that by contradiction that there exists $i$ such that
$\phi(i)$, $\phi(i+1)$ are ordered anti-clockwise on the circle.
Let $ i_0$ be the smallest integer such that $\phi(i_0)$, $\phi(i_0+1)$ are ordered anti-clockwise on the circle. 
By the assumption that $\phi(1)$, $\phi(2)$ are ordered 
clockwise on the circle, we know that $ i_0\geq 2.$ 
Since $\phi$ satisfies the above condition (1), it follows that  $\phi(i_0+1) = \phi(i_0-1).$ 
Let $\phi'$ be a new discrete path connecting $x$ to $y$ defined as follows:
$\phi'(i)=\phi(i)$ for $i\leq i_0-1$ and $\phi'(i)=\phi(i+2)$ for $i_0 \leq i \leq  k-2.$ Notice that the length of $\phi'$ is less than the length of $\phi$. This is a contradiction with the assumption that $\phi$ is a shortest discrete path connecting $x$ to $y$.

Now let $\phi$ be a shortest path 
 connecting $x$ to $y$ satisfying the above conditions (1) and (2).
 The discrete path $\phi$ have two possibilities, one going clockwise on the circle
and the other going anti-clockwise on the circle. The lengths of the two discrete paths connecting $x$ to $y$ in these two possibilities are respectively $|x-y|$ and $ n-|x-y|$.
Since $\phi$ be a shortest path connecting $x$ to $y$,  the length of $\phi$ is 
$$\min\{ |x-y|, n-|x-y|\}.$$
 \qed
 
$$\\$$

We have the following result, which gives a lower bound for the quantum complexity of any permutation.

\begin{theorem} Let $\beta$ be the permutation of the set $\{1, \cdots, n\}$.
If $$\beta=\rho_1 \cdots\rho_m$$ with each $\rho_i$ $(1\leq i\leq m)$ being a permutation from the generating set $\{\sigma,  \tau, \tau^{-1}\}$, then 
$$m\geq  \min \{|k-\beta(k)|, n-|k-\beta(k)|\} $$
for all $k\in \{1, \cdots, n\}$.

\end{theorem}

An interesting example is the transposition $(1~\lfloor \frac{n}{2}  \rfloor)$
switching $1$ with $ \frac{n}{2} $. By Theorem 2.3, this transposition has linear quantum complexity lower bound and by Lemma 4.1 it also has linear quantum complexity upper  bound.
 
$$\\$$

\noindent{Proof of Theorem 2.3}. Let $$\beta=\rho_1 \cdots\rho_m$$ with  each $\rho_i$ $(1\leq i\leq m)$  being a permutation from the generating set $\{\sigma,  \tau, \tau^{-1}\}$.

 We define a discrete path $\phi$  in $\{1, 2, \cdots, n\}$ as follows: $$\phi(1)=k, \phi(2) =\rho_m (k), \phi(3) =(\rho_{m-1}\rho_m) (k), \cdots, \phi(m+1) = (\rho_1 \cdots\rho_m)(k).$$ 

A crucial observation is that each permutation $\rho_i$ is an element from the generating set $\{\sigma, \tau, \tau^{-1}\}$ and hence it moves elements in $\{1, 2, \cdots, n\}$  by distance at most $1$, where the distance
is as defined in Definition 2.1.
It follows that $\phi$ is a discrete path as defined at the beginning of this section. 

By definition, we have 
$\phi(1) = k$ and  $\phi(m+1)=  \beta(k).$ This, together with Lemma 2.2,  implies that 
$$d(\phi(1), \phi(m+1)) \geq d(k, \beta(k)),$$
 where $d$ is the discrete path distance as in Definition 2.1. By Lemma 2.2, it follows that the length of the discrete path $\phi$ is
greater than or equal to  $  \min \{|k-\beta(k)|, n-|k-\beta(k)|\}.$
 By the definition of length of a discrete path, the length of $\phi$ is less than or equal to $m$. 
 
Combining the above two facts, we conclude that $$m\geq  \min \{|k-\beta(k)|, n-|k-\beta(k)|\}.$$ \qed

$$\\$$

 \section{Explicitly constructed permutations with  quadratic quantum complexity lower bound }\label{sec:neg}
 
 $$\\$$

In this section, we combine the discrete geometric method from Section 2 with  new combinatorial ideas  to give an explicit construction of permutations 
with  quadratic quantum complexity lower bound. We prove that very bumpy permutations have  quadratic quantum complexity lower bound.

$$\\$$
 
We need a few preparations. The following lemma plays an essential role in
the explicit construction of permutations 
with  quadratic quantum complexity lower bound.

\begin{lemma} Let $\omega$ be a permutation of $\{1, \cdots, n\}$ defined by
 $$\omega(1) =k_1, \cdots, \omega (n)= k_n.$$  We arrange the set
$$\{ d(1, k_1), d(2, k_2), \cdots, d(n, k_n)\}$$
 in a non-increasing order as
follows:
$$ d_1, d_2, \cdots, d_n, $$ 
where  $d$ is the discrete path distance in Definition 2.1.
 If 
$$\omega=(l_t ~l_t+1) \cdots (l_1~ l_1+1),$$ then for any $1\leq m\leq \frac{n}{2}$,
we have
$$t\geq d_1+d_2 +\cdots +d_{2m} - m^2.$$
\end{lemma} 

$$\\$$

\noindent{Proof of Lemma 3.1.} 
Let $$\omega_0=I, ~~~\omega_1=(l_1~l_1+1),~~~ \omega_2= (l_2~l_2+1) (l_1~l_1+1), ~~~ \cdots, \omega_t= (l_t~l_t+1)\cdots (l_1~l_1+1). $$

We identify the set $\{1, \cdots, n\}$ with the set of all integers modulo $n$.
Place the elements of  $\{1, \cdots, n\}$ on the circle in the complex plane by sending  $k\in \{1, \cdots, n\} $ to $exp(\frac{2\pi k}{n} ).$  The notation $k+l$ will 
be interpreted as $k+l$ modulo $n$ if $ k$ and $l$ are elements of $\{1, \cdots, n\}$.

For each $j\in \{1, 2, \cdots, n\}$,  if $k_j\neq j$, we define 
$$e_j(0)=0<e_j(1)< \cdots< e_j(s_j)$$ with $\omega_{e_j(s_j)}(j)= k_j$ as follows.

Let $e_j(1)$ to be the largest integer such that 
$$\omega_{e_j(1)-1}(j)=\omega_{e_j(0)}(j)=j$$ (note that  $\omega_{e_j (1)}(j)$ is either $j-1$ or $j+1$).

Inductively, for each $ i\geq 1$, if $\omega_{e_j(i-1)}(j)\neq k_j$,  
we define 
$e_j(i)$ to be the largest integer such that 
$$\omega_{e_j(i)-1}(j)=\omega_{e_j(i-1)}(j).$$

 Once $\omega_{e_j(i)}(j)=k_j$ for some $i$, we define 
this $i$ to be $s_j$.
The definition of $e_j (i)$ implies that  $\omega_{e_j (i)}(j)$ is either $\omega_{e_j(i)-1}(j)-1$ or $\omega_{e_j(i)-1}(j)+1$.
Hence either  $\omega_{e_j (i)-1}(j)=l_{e_j(i)}$ or $\omega_{e_j (i)-1}(j)=l_{e_j(i)}+1$.

By the definition of $ \{e_j(1), \cdots, e_j(s_j)\}$, we know that
 either the sequence $$(\omega_{e_j (0)}(j)=\omega_{e_j(1)-1}(j),\quad  \omega_{e_j(1)}(j)=\omega_{e_j (2)-1}(j), \cdots, \omega_{e_j(s_{j-1})}(j)=\omega_{e_j(s_j)-1}(j), \omega_{e_j(s_j)}(j) )$$ is  in anti-clockwise order on the circle:
$$\omega_{e_j (0)}(j)=\omega_{e_j(1)-1}(j)=j,   \omega_{e_j(1)}(j)=\omega_{e_j (2)-1}(j)=j+1, \cdots, $$
$$\omega_{e_j(s_j-1)}(j)=\omega_{e_j(s_j)-1}(j)=k_{j}-1,   \omega_{e_j(s_j)}(j)=k_{j}, $$ or 
the sequence $$(\omega_{e_j (0)}(j)=\omega_{e_j(1)-1}(j),\quad  \omega_{e_j(1)}(j)=\omega_{e_j (2)-1}(j), \cdots, \omega_{e_j(s_j-1)}(j)=\omega_{e_j(s_j)-1}(j), \omega_{e_j(s_j)}(j) )$$
is in clockwise order:
$$\omega_{e_j (0)}(j)=\omega_{e_j(1)-1}(j)=j, \omega_{e_j (2)-1}(j)= \omega_{e_j(1)}(j)=j-1, \cdots, $$
$$\omega_{e_j(s_j-1)}(j)= \omega_{e_j(s_j)-1}(j)=k_j +1, \omega_{e_j(s_j)}(j)=k_j.$$ 

In the anti-clockwise case, we have  
$$\omega_{e_j (i)-1}(j)= l_{e_j(i)}\quad \mbox{and} \quad
\omega_{e_j (i)}(j)= l_{e_j(i)}+1.$$
In the clockwise case, we obtain
$$\omega_{e_j (i)-1}(j)= l_{e_j(i)}+1\quad \mbox{and} \quad
\omega_{e_j (i)}(j)= l_{e_j(i)}.$$

This implies that $s_j$ is either $|k_j-j|$ or $n-|k_j-j|.$
Hence $s_j\geq d(j, k_j).$

Let $$ E(j) = \{ e_j (1), e_j(2), \cdots, e_j(s_j)     \}.$$

If $k_j=j$, we define $E(j)$ to be the empty set $\emptyset$.

Define a discrete path $\phi_j$ in $\{1, 2, \cdots, n\}$ by:
$$\phi_j(1)=\omega_{0} (j)=j, \phi_j (2)=\omega_{e_j(1)}( j), \phi_j(3) =
\omega_{e_j(2)} (j), \cdots, 
\phi_j (s_j+1)=\omega_{e_j(s_j)}(j)=k_{j}. $$

It is easy to see that $\phi_j$ is a discrete path as defined in Section 2.

$$\\$$

\noindent{\bf Claim 1:} Let $j\neq j'$ be a pair of integers in $\{1, 2, \cdots, n\}$.

 (1) If the two discrete paths $\phi_j$ and $\phi_{j'}$ travel in the same direction on the circle (either clockwise or anti-clockwise), then 
 $E(j)\cap E(j')=\emptyset$;
 
 (2) If the two discrete paths $\phi_j$ and $\phi_{j'}$ go in opposite direction on the circle, then the number of elements in $E(j)\cap E(j')$ is at most two.
 
$$\\$$

\noindent{Proof of Claim 1.} 
We first prove part (1). Without loss of generality,
we can assume that both discrete paths $\phi_j$ and $\phi_{j'}$ travel in the anti-clockwise direction on the circle.

Assume by contradiction  $E(j)\cap E(j')\neq \emptyset$. This implies that there exist $k$ and $k'$ satisfying $e_j(k)=e_{j'}(k')$.
Hence we have 
$$(l_{e_j(k)}, l_{e_j(k)}+1) = (l_{e_{j'}(k')}, l_{e_{j'}(k')}+1).$$
By the assumption that  both discrete paths $\phi_j$ and $\phi_{j'}$ travel in the anti-clockwise direction on the circle, we know 
$$ \omega_{e_j (k)} (j )=l_{e_j(k)}+1, \quad \quad \quad  \omega_{e_{j'}(k')}(j') =l_{e_{j'} (k')}+1.$$
As a consequence, we have
$$\omega_q(j)=\omega_q(j')$$ for  $q=e_j (k)=e_{j'}(k').$

The above equation implies  $j=j'$, a contradiction with the assumption that $j\neq j'$.
This completes the proof of part (1) of the Claim 1.

Now we prove part (2) of the Claim 1. Without loss of generality, we assume 
that $\phi_j$  travels in the anti-clockwise direction on the circle
and $\phi_{j'}$  travels in the clockwise direction on the circle.

In this case, we have 
$$\omega_{e_{j'}(1)-1}(j')=j', \quad \omega_{e_{j'}(1)}(j')=j'-1,\quad \omega_{e_{j'}(2)-1}(j')=j'-1, \quad\omega_{e_{j'}(2)}(j')=j'-2 ,$$ $$ \cdots, 
\quad \omega_{e_{j'}(s_{j'})-1}(j')=k_{j'}+1, \quad \omega_{e_{j'}(s_{j'})}(j')=k_{j'}. $$

Let $k$ and $k'$ be a pair of integers such that $e_k (k)= e_{k'} (k')$.
This assumption implies 
$$(l_{e_j(k)}, l_{e_j (k)}+1) = (l_{e_{j'}(k')}, l_{e_{j'}(k')}+1).$$
By the other assumption that $\phi_j$  travels in the anti-clockwise direction on the circle
and $\phi_{j'}$  travels in the clockwise direction on the circle, we have
$$\omega_{e_j(k)} (j )=l_{e_j(k)}+1, \quad \quad \quad \omega_{e_{j'}(k')}(j')=l_{e_{j'} (k')} . $$
As a consequence, we obtain
$$    \omega_{q}(j)+1=\omega_{q}(j')   $$
for $q=e_j(k)$. Since the two  discrete paths $\phi_j$ and $\phi_{j'}$ travel in opposite direction on the circle   and each of them goes around the circle at most once, there are at most two integers $q$ in
$$\{e_j(0), e_j (1), \cdots, e_j( s_j) \} \cap \{e_{j'}(0), e_{j'} (1), \cdots, e_{j'}( s_{j'}) \}$$
satisfying  
$$\omega_{q}(j)+1=\omega_{q}(j').$$
It follows that there are at most two pairs of $k$ and $k'$ satisfying $e_j (k)= e_{j'} (k')$. This completes the proof of part (2) of the Claim 1.

$$\\$$

Recall $ d(j, k_j)\leq s_j$ from the discussions before Claim 1.
For each $j\in\{1,2, \cdots, n\}$, let  $F(j)$ be the subset of $E(j)$ consisting of the first $d(j, k_j)$ number of elements in the sequence $e_j (1), e_j(2), \cdots, e_j(s_j).$  The following claim is a variation of Claim 1.
We mention that Claim 1 already suffices to obtain a weaker version of Lemma 3.1, which can be used to prove quadratic quantum complexity lower bound for the permutation in Theorem 1.1 with a smaller coefficient of $n^2$.

$$\\$$

\noindent{\bf Claim 2:} For each $j$, let $\psi_j$ of the discrete path
obtained by restricting $\phi_j$ to the domain $\{1,2, \cdots, d(j, k_j)\}.$
Let $j\neq j'$ be a pair of elements  $\{1, 2, \cdots, n\}$.

(1) If the two discrete paths $\psi_j$ and $\psi_{j'}$ go in the same direction on the circle (either clockwise or anti-clockwise), then 
 $F(j)\cap F(j')=\emptyset$;

(2) If the two discrete paths $\psi_j$ and $\psi_{j'}$ go in opposite direction on the circle, then the number of elements in $F(j)\cap F(j')$ is at most one.

$$\\$$

\noindent{Proof of Claim 2.} Part (1) of Claim 2 follows from part (1) of Claim 1. 

Part (2) of Claim 2 can be proved using essentially the same argument as
in the proof of part (2) of Claim 1. Without loss of generality, we assume 
that $\psi_j$  travels in the anti-clockwise direction on the circle
and $\psi_{j'}$  travels in the clockwise direction on the circle.
By the definitions of $F_j$ and $F_{j'}$, we know   that the discrete paths $\psi_j$ and $\psi_{j'}$ travel at most half of the circle since $d(j, k_j)\leq \frac{n}{2}$ and $d(j', k_{j'})\leq \frac{n}{2}.$ Hence there is at most one integer $q$  in 
$$\{e_j(0), e_j (1), \cdots, e_j(d(j, k_j))\} \cap \{e_{j'}(0), e_{j'} (1), \cdots, e_{j'}(d(j', k_{j'})) \}$$
satisfying  
$$\omega_{q}(j)+1=\omega_{q}(j').$$
It follows that there is at most one pair 
of $k\leq d(j, k_j)$ and $k'\leq d(j', k_{j'})$ satisfying $e_j (k)= e_{j'} (k')$. This  completes the proof for part (2) of  Claim 2. 

$$\\$$

Let $j_1, j_2, \cdots, j_{2m}$ be integers in $\{1, 2, \cdots, n\}$
such that $$d(j_1, k_{j_1})=d_1, d(j_2, k_{j_2})=d_2, \cdots, d(j_{2m}, k_{j_{2m}}) =d_{2m}.$$
We define $A$ (respectively  $B$) to be the set consisting of $j_i$ $(1\leq i\leq 2m)$ whose  corresponding  discrete path $\phi_{j_i}$ 
is in anti-clockwise (respectively clockwise) order  on the circle.

Let $a$ (respectively $b$) be the number of elements in $A$ (respectively $B$).
We have $$ a+b=2m.$$
Hence $$ab\leq \left(\frac{a+b}{2}\right)^2=m^2.$$

By Claim 2 and  Lemma 2.2, it follows that

$$ \# (      F(j_1)\cup  F(j_2) \cup \cdots \cup F(j_{2m}) )= \# ( (\cup_{i\in A} F(i) ) \cup (\cup_{j\in B} F(j)) ) 
$$
$$\geq (d_1+d_2+\cdots + d_{2m})- ab\geq (d_1+d_2+\cdots + d_{2m})-m^2,$$
where the notation $\#$ denotes the number of elements in the set.
The above inequality implies Lemma 3.1.
\qed

$$\\$$

\begin{lemma} Let $\omega$ be a permutation  of $\{1, \cdots, n\}.$
If we can write $$\omega=\rho_1 \cdots\rho_t$$ with each $\rho_j$ $(1\leq j\leq t)$ being a permutation from the generating set $\{\sigma, \tau, \tau^{-1}\}$,  then there exists $k$ such that
$$\tau^{k}\omega=(l_i ~l_i+1) \cdots (l_1~ l_1+1)$$ 
with $$t\geq 2i-1.$$

\end{lemma}

$$\\$$

\noindent{Proof of Lemma 3.2.} 
By assumption, we can write
$$\omega=\tau^{t_0}\sigma\tau^{t_1}\cdots \tau^{t_{i-1}} \sigma \tau^{t_i}$$ such that $|t_0|+|t_1|+\cdots +|t_i|+i=t$, where  $t_0$ and $t_i$ may be $0$ and $t_j\neq 0$ for any $1<j<i$.

We rewrite
$$\omega=\tau^{t_0+t_1+\cdots t_i}(\tau^{-(t_1+\cdots+t_i)}\sigma\tau^{t_1+\cdots+t_i})\cdots (\tau^{-(t_{i-1}+t_i)}\sigma\tau^{t_{i-1}+t_i}) (\tau^{-t_i}\sigma \tau^{t_i}).$$

Note that $\tau^{-(t_1+\cdots+t_i)}\sigma\tau^{t_1+\cdots+t_i}, \cdots, \tau^{-(t_{i-1}+t_i)}\sigma\tau^{t_{i-1}+t_i}, $ and $\tau^{-t_i}\sigma \tau^{t_i}$ are
all adjacent transpositions. The inequality $t=|t_0|+|t_1|+\cdots +|t_i|+i \geq 2i-1$ follows from the fact that $|t_j|\geq 1$ for all $1\leq j\leq i-1.$ 
Lemma 3.2 now follows.\qed

$$\\$$

The following lemma describes the bumpiness of the permutation $\omega$ in Theorem 1.1. This bumpiness  is the reason for the permutation to have quadratic quantum complexity lower bound (c.f. Definition 1.1, Theorem 1.2
and Remark 5.3).

$$\\$$

\begin{lemma} Let $\omega$ be the  permutation 
$\omega$ of $\{1, \cdots, n\}$ in Theorem 1.1.
For any $0\leq k\leq n-1$, if we write
$$(\tau^k\omega) (1) =k_1, \cdots, (\tau^k\omega)(n)=k_n,$$
then $$d_1 \geq \frac{n}{2}-1, d_2 \geq \frac{n}{2} -1, d_3\geq \frac{n}{2}-2, d_4\geq \frac{n}{2}-2, \cdots,d_{2m-1}\geq \frac{n}{2}-m, d_{2m}\geq \frac{n}{2}-m$$
for any $1\leq m\leq \frac{n}{2},$
 where $d$ is the discrete path distance in Definition 2.1 and $$\{ d(1, k_1), d(2, k_2), \cdots, d(n, k_n)\}$$ is rearranged 
 in the following non-increasing order:
$$ d_1, d_2, \cdots, d_n. $$

\end{lemma}

$$\\$$

\noindent{Proof of Lemma 3.3.} We identify the set $\{1, 2, \cdots, n\}$ with the set of all integers modulo $n$.
When $n$ is odd, by the choices of $\omega(1), \cdots, \omega(n)$ in Theorem 1.1, we have 
$$\{ \omega(1)-1, \omega(2)-2, \cdots,\omega(n)-n\}=\{0, 1, 2, \cdots, n-1\},$$
where the subtraction is done modulo $n$.
It follows that, for any integer $k$, we have 
$$\{ k+ \omega(1)-1, k+\omega(2)-2, \cdots, k+\omega(n)-n\}=\{0, 1, 2, \cdots, n-1\},$$
where the addition and subtraction are done modulo $n$.
The above equality implies that
$$\{d(1, k+\omega(1)), \quad d(2, k+\omega(2)),\quad \cdots,\quad d(n, k+\omega(n))\}$$
$$= \{0,  1,  2,  \cdots, \ \frac{n-1}{2},  1, \ 2,  \cdots,  \frac{n-1}{2}\} \quad \quad (\mbox{counted with multiplicity}).$$
This completes the proof of Lemma 3.3 when $n$ is odd. 
 The even case is similar. \qed

Now we are ready to prove Theorem 1.1.

$$\\$$

\noindent{Proof of Theorem 1.1.} 
Let $\omega$ be the permutation in Lemma 3.3.
If we can write $$\omega=\rho_1 \cdots\rho_t$$ with each $\rho_j$ $(1\leq j\leq t)$ being a permutation from the generating set $\{\sigma, \tau, \tau^{-1}\}$,
then by Lemma 3.2, we can write 
$$\tau^k \omega= (l_i ~l_i+1) \cdots (l_1~ l_1+1)$$ for some integer $k$
such that $t\geq 2i-1$.

Choose $m=\lfloor \frac{n}{4} \rfloor.$
By Lemma 3.3, we have 
$$d_1\geq \frac{n}{2}-1, d_2\geq \frac{n}{2}-1, d_3\geq \frac{n}{2}-2, 
d_4\geq \frac{n}{2}-2, \cdots, d_{2m-1}\geq \frac{n}{2}-m, d_{2m}\geq \frac{n}{2}-m.$$
Applying Lemma 3.1 to $\tau^{k}\omega$, we obtain 
$$i\geq (d_1+\cdots d_{2m}) -m^2 \geq 2 ( (\frac{n}{2}-1)+\cdots + ( \frac{n}{2}-m)) -m^2 .$$ Hence we have
$$i\geq m(n-2m-1)\geq \frac{n^2-2n-3}{8}.$$
By Lemma 3.2, it follows that $$t\geq 2i-1\geq  \frac{n^2-2n-7}{4}.$$ 
This completes the proof of Theorem 1.1.
\qed

$$\\$$

The following is an example of a permutation of $\{1, 2, \cdots, n\}$ which 
satisfies the conclusion of Lemma 3.3 for $k=0$ but still can be implemented linearly. This example shows that in the proof of Theorem 1.1, it is essential for  Lemma 3.3 to hold for all $0\leq k \leq n-1$.

$$\\$$

\begin{example}
Let $n=2j$. Define a permutation $\beta$ of $\{1, 2, \cdots, n\}$ by:
$$\beta(1) = j+1, \quad \beta(2)= j+2, \cdots, \beta(j)=2j, \quad \beta(j+1) =1, \dots, \quad \beta(n)=j.$$  Note $\beta=\tau^j$. Hence the permutation $\tau$ can be implemented linearly.

\end{example}

$$\\$$

The proof of Theorem 1.3 is similar to that of Theorem 1.1.

$$\\$$

\noindent{Proof of Theorem 1.3.} We follow the strategy of the proof for  Theorem 1.1. 

Choose $m=\left \lfloor \frac{n}{8} \right\rfloor,$ 
the floor of $\frac{n}{8}.$ Let $d_1, d_2, \cdots, d_{2m}$ be as in Lemma 3.1. 

By Definition 1.2, we have
$$d_1+d_2+\cdots + d_{2m}-m^2\geq  \frac{n}{8} \left(2  \left\lfloor \frac{n}{8} \right\rfloor\right)- \left\lfloor \frac{n}{8} \right\rfloor^2 $$
$$ = \left(\frac{n}{4} -  \left\lfloor \frac{n}{8} \right\rfloor   \right) \left\lfloor \frac{n}{8} \right \rfloor 
= \left(\frac{n}{8}+( \frac{n}{8}-  \left\lfloor \frac{n}{8} \right\rfloor )  \right) \left(\frac{n}{8}-( \frac{n}{8}-  \left\lfloor \frac{n}{8} \right\rfloor )  \right)   $$
$$= \left(\frac{n}{8}\right)^2-\left( \frac{n}{8}-  \left\lfloor \frac{n}{8} \right\rfloor  \right) ^2
\geq  \frac{n^2}{64}-1. $$

By Lemma 3.1 and Lemma 3.2, we obtain $$l\geq  2\left(\frac{n^2}{64}-1\right)-1\geq  \frac{n^2}{32}-3.$$ 
 \qed

$$\\$$

\section{All permutations have quadratic quantum complexity upper bound}\label{sec:pre}

$$\\$$

In this section, we prove that all permutations in symmetric groups $S_n$ have
quadratic quantum complexity upper bound (Theorem 1.4). 

We need some preparations to prove Theorem 1.4. 
Recall that  the transposition $(k~l)$  is the permutation switching $k$ with $l$ and leaves every other element in the set $\{1,2,\cdots, n\}$ unchanged.

$$\\$$

\begin{lemma} For each positive integer  $l\leq n$, the transposition $(1~l )$ 
 can be implemented linearly, more precisely,  if we write 
$$(1~l ) =\rho_1 \cdots\rho_m$$
with each $\rho_i$ $(1\leq i\leq m)$ being a permutation from the generating set $\{\sigma,  \tau, \tau^{-1}\}$ and 
$$m\leq 2n-3.$$
\end{lemma}

$$\\$$

\noindent{Proof of Lemma 4.1.} We have

$$(1~l)= (1~2) (2~3)\cdots (l-2~l-1) (l-1~ l) (l-2~l-1) (l-3~l-2) \cdots (2~3) (1~2).$$

Notice that 
$$ (2~3)= \tau \sigma \tau^{-1}, \cdots, (l-2~l-1)=\tau^{l-3}\sigma \tau^{-(l-3)}, (l-1~l)= \tau^{(l-2)}\sigma \tau^{-(l-2)}.$$

Plugging these equations into the previous identity, after certain (magical) cancellations of $\tau$ and $\tau^{-1}$,  we obtain

$$(1~l)= \underbrace{ \sigma \tau \sigma \tau \sigma \cdots \tau \sigma \tau \sigma\tau^{-1} \sigma\tau^{-1}\sigma \tau^{-1} \cdots \sigma \tau^{-1} \sigma}_{\text 4l-7} .$$

Note that there are a total of $4l-7$ terms on the right hand side of the above equation. 

We can now apply the same argument using the clockwise route from $1$ to $l$:

\noindent $1, n, n-1, \cdots, l$ (instead of the  above anti-clockwise route:  $1, 2, \cdots, l$):
$$(1~l)=(1~n)(n~n-1)\cdots (l+2~l+1)(l+1~l)(l+1~l+2)\cdots (n-1~n) (n~1).$$
 This way, we can write $(1~l)$ as a product of at most $4(n-l)+3$ number of terms  from  the generating set $\{\sigma,  \tau, \tau^{-1}\}$. 

Summarizing the above discussions, we can write 
$(1~l)$ as a product of at most $\min\{4l-7,  4(n-l)+3\}$ number of terms  from  the generating set $\{\sigma,  \tau, \tau^{-1}\}$. 
From the equation 
$$(4l-7)+  (4(n-l)+3)=4n-4,$$ we obtain
$$\min\{4l-7,  4(n-l)+3 \}\leq 2n-2.$$
Since both $ 4l-7$ and $  4(n-l)+3$ are odd number,
we have $$\min\{4l-7,  4(n-l)+3\}\leq 2n-3.$$
This completes the proof of Lemma 4.1.
 \qed
 
 $$\\$$
We are now ready to prove Theorem 1.4.

$$\\$$

\noindent{Proof of Theorem 1.4.}   
Without loss of generality, we assume that $l\geq k$.
We can easily verify the formula $$ (k~l)=   \tau^{k-1} ( 1~ l-(k-1))  \tau^{-(k-1)}.$$
The equation $\tau^{k-1}= \tau^{n-(k-1)}$ implies that $ \tau^{k-1}$ can be implemented using at most $\frac{n}{2}$ elements from the generating set
$\{ \sigma, \tau, \tau^{-1}\}$.
Hence by Lemma 4.1
 any transposition
$(k~l)$ can be implemented by $ \frac{n}{2} + (2n-3) +\frac{n}{2} = 3(n-1)  $  elements from the generating set $\{\sigma,  \tau, \tau^{-1}\}$.

We recall that a cycle $(x_1~ x_2~\cdots ~x_k)$ is a permutation
that sends $x_j$ to $x_{j+1}$ for all $1\leq j \leq k-1$, sends $x_k$ to
$x_1$, and keeps all other elements in $\{1, 2, \cdots, n\}$ unchanged.
We define the length of the cycle $(x_1~ x_2~\cdots~ x_k)$  to be $k$.
Each permutation $\omega$ in $S_n$ is a product of cycles $\gamma_1, \cdots, \gamma_q$:
$$\omega= \gamma_1 \gamma_2\cdots \gamma_q$$
such that $$l_1 +l_2+\cdots + l_q\leq n, $$
where $l_i$ is the length of the cycle $\gamma_i$ for all $1\leq i \leq q$.

The cycle $\gamma_i$ can be expressed as the product of transpositions as follows:

$$\gamma_i= (x_1~x_2~\cdots~ x_{l_i})=(x_1~x_2) (x_2~x_3) \cdots (x_{l_i-1}~x_{l_i}),$$
where $1\leq i\leq q$. This implies that $\gamma_i$ can be implemented by 
at most $3(n-1)(l_i-1))$ number of elements from the generating set $\{\sigma,  \tau, \tau^{-1}\}$. 

Combining the above facts, we conclude that the number of elements $\omega$ 
needed from the generating set $\{\sigma,  \tau, \tau^{-1}\}$ is 
 at most 
$$3(n-1)( l_1 +l_2+\cdots + l_q -q)\leq 3(n-1)^2.$$
Hence $\omega$   can be quadratically   implemented. \qed

$$\\$$

\section{Almost all permutations have quadratic quantum complexity lower bound}\label{sec:neg}

$$\\$$

In this section, we prove that almost all permutations have quadratic quantum complexity lower bound.  

$$\\$$

\noindent{Proof of Theorem 1.5.} We prove Theorem 1.5 by showing that almost all permutations are very bumpy.

For each $k\in \{0, 1, 2, \cdots, n-1\}$, we first count the number of permutations which do not satisfy the   inequality in Definition 1.2:

$$\# \{\quad i \in \{1, 2, \cdots, n\}: d(k+ \omega(i), i )\geq \frac{n}{8}\quad \}
\geq  \frac{n}{4} \quad \quad \quad \quad \quad \quad ~~~~~~~~~~~~(\ast). $$

If a permutation $\omega$ does not satisfy the above condition $(\ast)$, 
then there exist at least $p=n-\lceil \frac{n}{4}\rceil$ number of $i\in \{0, 1, 2, \cdots, n-1\}$
such that 
$$    d(k+ \omega(i), i ) <\frac{n}{8} \quad \quad \quad \quad \quad \quad ~~~~~~~~~~~~(\ast\ast) .$$

Given $k\in \{0, 1, 2, \cdots, n-1\}$, for each subset $B$  of $\{ 1, 2, \cdots, n\}$ with $\# B=p$, 
the number of permutations $\omega$ satisfying condition $( \ast\ast)$
for all $i\in B$ is at most
$$\left(  \frac{n}{4} +1\right) ^{ p }
 (n- p )! . $$
  This is because for each $i\in B$, $\omega(i)$ has at most $ \frac{n}{4} +1$ possible choices of values satisfying $(\ast \ast)$,
  and once we fix the values of $\omega$ on $B$, there are $(n-p)!$ possibilities of choosing the permutations $\omega$.

  We also have $ {n \choose p } $ ways of choosing subset $B$ of $\{ 1, 2, \cdots, n\}$  satisfying the condition $\# B=p.$ 
  
In summary, for each $k\in \{0, 1, 2, \cdots, n-1\}$,  the total number of  permutations  $\omega$ satisfying the above condition $(\ast\ast)$ for at least $p$ number of $i$ is at most

$$ {n \choose p } 
 \left(  \frac{n}{4} +1\right) ^{ p }
 (n- p) ! .  $$

Hence the total number of not very bumpy permutations is at most

$$ n{n \choose p}  \left(  \frac{n+4}{4} \right) ^{ p } (n-p)!.  $$

Using calculus,  we have the following estimate
$$\mbox{ln} (p!)=\mbox{ln}2+\cdots+\mbox{ln}p\geq  \int_{1}^{p} \mbox{ln}x dx
$$
$$= ( x\mbox{ln}x-x)|^p_1=  p\mbox{ln}p-(p-1).$$

It follows that $$ p! \geq \frac{p^p}{e^{p-1}}. $$

Applying this formula, we obtain
$$ \frac{1}{n!} n {n \choose p}  \left(  \frac{n+4}{4} \right) ^{p} (n-p)! = 
\frac{n}{p!}  \left(  \frac{n+4}{4} \right) ^{ p }
 \leq  n \frac{e^{ p-1}}{ p^p}  \left(  \frac{n+4}{4} \right) ^{p}$$
$$= e^{-1}n \left(\frac{e(n+4)}{4p} \right)^p 
< n \left(\frac{e(n+4)}{4 ( \frac{3n}{4}-1)} \right)^p  =n \left(\frac{e(1+\frac{4}{n})}{3-  \frac{4}{n}} \right)^p $$
$$< n \left(\frac{e(1 +\frac{1}{10})}{2.999} \right)^p \leq n \left( \frac{2.992 }{2.999} \right)^{ \frac{3n}{4} -1}  \longrightarrow 0$$ 
as $n\rightarrow \infty$, where we use the inequality $ \frac{3n}{4} -1 \leq p \leq  \frac{3n}{4} $ and $e<2.72$ in the above estimate and  assume that
$n\geq 4000$.

By Theorem 1.3, this completes the proof of Theorem 1.5.
\qed

$$\\$$

\noindent {\bf Remark} 5.3.
More generally, for any $0<b<\frac{1}{2}$ and $0<c<\mbox{min}\{4b,1\}$, we can define
a permutation $\omega$ of  $\{1, 2, \cdots, n\}$
 to be $(b,c)$-bumpy if 
  $$\# \{\quad i \in \{1, 2, \cdots, n\}: \quad d(k+ \omega(i), i )\geq bn \quad \}
\geq  cn$$
for all $k\in \{0, 1,2 \cdots, n-1\}. $  
The same method can be used to prove that  any $(b,c)$-bumpy permutation has quadratic quantum complexity lower bound (with $\frac{1}{2}(4bc-c^2)$ as the coefficient of $n^2$).
If in addition $2 eb +c<1$, then the probability for permutations to be $(b, c)$-bumpy goes to $1$ as $n\rightarrow \infty$.

In fact, for all positive number $c_0<\frac{1}{ 2e(e+2)}$, we can find positive constants
$b$ and $c$ satisfying the above conditions such that 
any $(b,c)$-bumpy permutation has quadratic quantum complexity lower bound with $c_0$ as the coefficient of $n^2$
and the probability for permutations to be $(b, c)$-bumpy goes to $1$ as $n\rightarrow \infty$.

Note that in Theorem 1.3 and  Theorem 1.5, the constants are chosen to be
$b=\frac{1}{8}$ and $c=\frac{1}{4}.$

$$\\$$

\section{Concluding Remarks}\label{sec:neg}

$$\\$$

In this article, we essentially obtain the optimal results on complexity upper
bound for all permutations and lower bound for very bumpy permutations
as both upper and lower bounds are quadratic.
The method used in this article can potentially be used to study quantum complexity for more general unitaries in quantum computations.
In the future, we plan to investigates the complexity problem for 
permutations on the set of $n$-strings of two bits with respect to local logic gates.

$$\\$$

\noindent {\bf Acknowledgement:} I would like to express my gratitude to Professor Arthur Jaffe  for offering me the opportunity to conduct research in his research  group at Harvard University and for being a wonderful mentor. I  am deeply grateful to  Professor Michael Freedman for his kind advice and for suggesting  the fascinating problem of constructing explicit  hard to implement permutations to me at the conference ``New Frontiers: Interactions between Quantum Physics and Mathematics". I  would like to thank 
Dr. Kaifeng Bu for teaching me quantum computations and for many stimulating discussions. I   also wish to  thank Professor Zhenghan Wang and Professor Scott Aaronson for very helpful discussions.

$$\\$$

\end{document}